# Towards Improved Polarization Uniformity in Ferroelectric $Hf_{0.5}Zr_{0.5}O_2$ Devices within Back End of Line Thermal Budget for Memory and Neuromorphic Applications

Padma Srivari, Ella Paasio, Xinye Li, Sayani Majumdar
*Information Technology and Communication Sciences, Tampere University, FI – 33720 Tampere, Finland*
E-mail: sayani.majumdar@tuni.fi

**Abstract —** Thin film ferroelectric devices with ultralow power operation, non-volatile data retention and fast and reliable switching are attractive for non-volatile memory and as synaptic weight elements. However, low thermal budget ferroelectric oxides suffer from crystalline inhomogeneity and defects that makes their large-scale circuit integration challenging. Here, we report on the thermally engineered way to induce wafer-scale homogeneity in $Hf_{0.5}Zr_{0.5}O_2$ capacitors that can lead to high device reliability making their integration possible in ultralow power memory and neuromorphic computing hardware.

## INTRODUCTION

Today's data-centric computing requires dense integration of memory and logic components for high-performance, reliable and fast computing operations under low power requirement and heat dissipation [1]. Neuromorphic computing additionally requires physical proximity of memory and logic units to minimize data transfer time and energy and multi-level operation of memory elements to enable on-chip synaptic weight storage and updates through hardware multiply-and-accumulate (MAC) operation. This necessitates 3D vertical stacking of memory devices with logic circuits and requires compatibility of processes for the logic and memory components. To meet these multitude of requirements, device scaling, yield, reproducibility and reliability is of vital importance. The semiconductor industry is currently at a juncture where physical downsizing of gate length of complementary metal oxide semiconductor (CMOS) transistors is no longer possible. Hence, different novel device and architecture level options are under active investigation.

Ferroelectric (FE) components are one of the strong contenders for non-volatile memory (NVM), synaptic weight elements and in-memory logic components due to their voltage-driven low power operation, fast and multi-level switching, long data retention and high endurance [2]. In recent years, CMOS compatible Hafnia-based ferroelectric materials gained high research interest due to properties like high remnant polarization ($P_r$), scalability, possibility of integration with CMOS front-end (FEOL) or back-end-of-line (BEOL) processes and so on, making them suitable for mainstream semiconductor industry [3]. However, like other emerging technologies, FE technology still suffers from different non-idealities like low yield, reliability, device-to-device or cycle-to-cycle variability etc. In this work, we report on material challenges that bring device-level yield, uniformity and reproducibility issues due to BEOL compatible low thermal budget process and demonstrate possible mitigation strategies that could make their integration possible in 1T-1C ferroelectric random-access memory (FeRAM) array or ferroelectric field effect transistor (FeFET) based memory and synaptic circuits.

## EXPERIMENTAL

The Metal-FE-Insulator-Metal (MFIM) capacitors were fabricated using Atomic Layer Deposition (ALD) and metal evaporation. The bottom TiN (30 nm) electrode was grown by plasma-enhanced ALD (PEALD) while the $Hf_{0.5}Zr_{0.5}O_2$ (HZO) (10 nm) and $Al_2O_3$ (1.2 nm) layer was fabricated using thermal ALD. 50:50 ratio of the Hf:Zr was obtained using alternate cycles of the Hf and Zr precursors [4]. Following the HZO deposition, $Al_2O_3$ capping layer was deposited without breaking vacuum. Growth temperature of the HZO and $Al_2O_3$ layer was conducted at 200 $^0$C and 285 $^0$C. After the film deposition, rapid thermal annealing (RTA) step was done for 30 seconds at annealing temperature of 450 °C, 550 °C and 600 °C under nitrogen atmosphere (samples named S200-450, S200-550, S200-600 and S285-450, S285-550 and S285-600, respectively). Finally, the FE capacitor stack was completed by evaporating Ti/Au top electrode through electron beam evaporation. Electrical characterizations of the samples were carried out by applying voltage pulses between the bottom TiN and the top gold electrode. For ferroelectric polarisation - voltage ($P$–$V$) characteristics, and endurance measurements, ferroelectric material tester aixACCT 2000E was used.

**RESULTS AND DISCUSSION**

*A. Experimental*

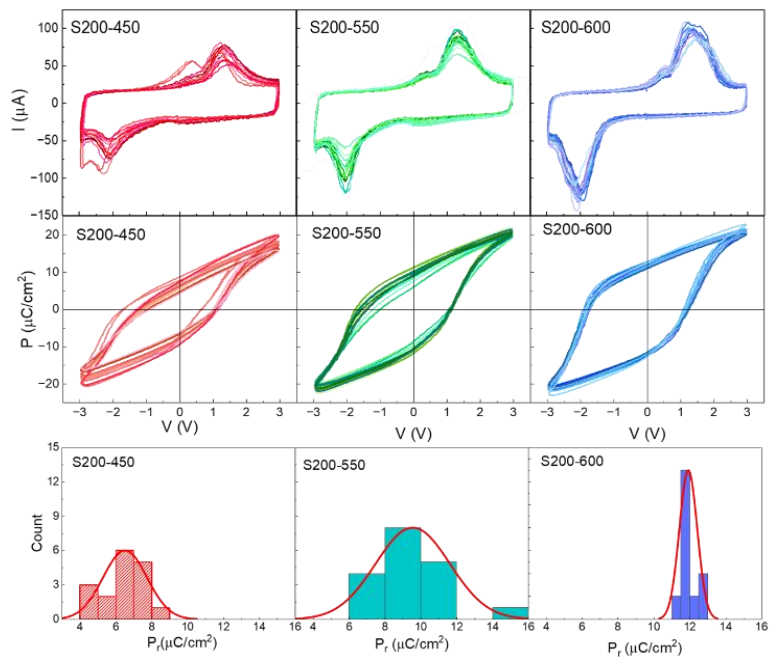

Fig. 1. *I-V* and *P-V* plots measured with dynamic hysteresis measurement (DHM) pulsing scheme of 1 kHz frequency and distribution of $P_r$ values for the devices grown at 200 °C showing wider distribution of parameters for samples annealed at lower temperature due to smaller ferroelectric grain size.

50% Zr-doped Hafnia, commonly known as HZO, is the most studied FE material for BEOL integration compatible memory elements. There are multitudes of reports on FE performance due to different deposition and annealing conditions, however, their impact on the wafer-scale uniformity of properties, reliability, fatigue and breakdown is not often discussed which make the assessment impossible for their large-scale circuit implementation.

In Fig. 1, we present the distribution of device performances from one-quarter of a 150 mm wafer by studying the dynamic current – voltage ($I$ – $V$) and *P-V* data from samples grown at 200 °C (S200-450, S200-550, S200-600). From the device-to-device (D2D) variation perspective, the sample annealed at 600 °C showed the best performance indicating highest $P_r$ and D2D reproducibility. For the samples annealed at 450 °C and 550 °C, the $P_r$ value is low and a significant D2D variation exists that can be attributed to incomplete crystallization to FE orthorhombic phase and spatial inhomogeneity in the HZO film caused by the smaller FE grain size that was also previously observed in low thermal budget HZO films [5, 6].

This result, together with results from De et al. [5] gives a fairly good idea that to improve D2D variation, a higher annealing temperature or longer annealing duration is important. However, annealing temperature of 600 °C is too high for the BEOL processes where already degradation of FEOL transistors and metal interconnect layers start to happen. Therefore, alternate strategies for improvement of $P_r$ and wafer-scale uniformity are needed.

In a recent work, Lee et al. [7] demonstrated larger crystallite grain size and superior FE properties can be obtained by increasing the HZO deposition temperature. Also, in our previous experiments [8], we found that ALD grown $Al_2O_3$ layer gives improved leakage performance when grown at 300 $^0$C. To test this hypothesis for improved electrical performance reliability, we tested samples grown at 285 $^0$C and annealed at the same temperatures as above. Fig. 2 shows D2D variations for the sample grown at 285 $^0$C and annealed at 450 $^0$C and 550 $^0$C (S285-450 and S285-550, respectively). The samples showed nearly double FE polarization compared to the S200 samples with moderate D2D variation confirming better crystallinity of the samples grown at higher temperatures. It was found that upon a few electric field cycling, the $P_r$ value improved even further with distribution of the switching voltage narrowing significantly. The data in fig. 2 are taken after 6 field cycling. Here, it is important to mention that the increased $P_r$ value in high temperature grown or annealed films comes at a price of increased leakage current and limited endurance in HZO films, that was reported in our previous works [6] and by others [9, 10].

*B. Modelling*

To clarify the role of thermal processing-induced different microcrystalline properties of the 10 nm HZO films and its impact on the FE polarization and leakage currents, we performed ferroelectric Jiles-Atherton (JA) modelling on the two sets of HZO device (S200 and S285) data.

In our previous work, we have shown the suitability of the JA model for reproducing the major and minor loop operations of the HZO capacitors [11]. Here, we investigate the role of different microcrystalline properties on the JA model parameters.

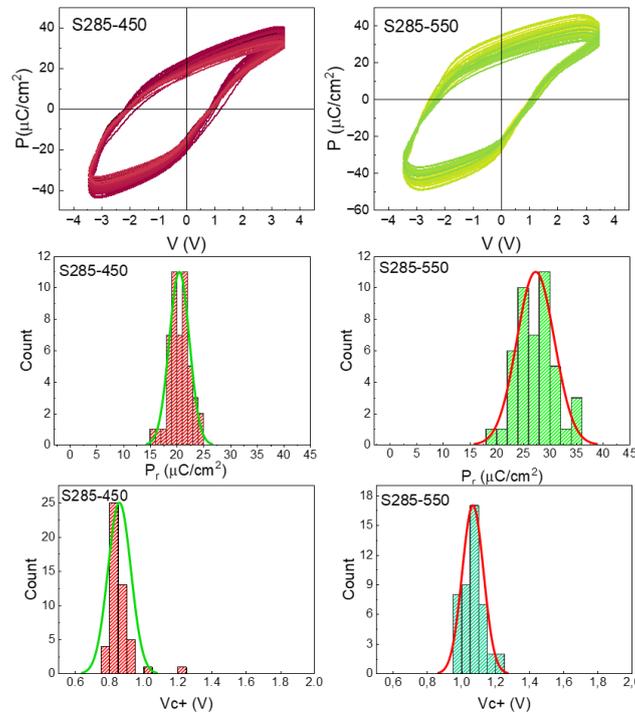

Fig. 2. *P-V* loops from multiple different devices from S285-450 and S285-550 samples after 6 cycles of wakeup showing higher $P_r$ and a narrower distribution of $P_r$ and $V_C$ values for the devices grown at higher deposition temperatures.

In the JA model, the polarization is a combination of reversible and irreversible polarization ($P_{\text{irr}}$) contributions, which are considered through the anhysteretic polarization ($P_{\text{anh}}$)

$$P = (1-c)P_{\text{irr}} + cP_{\text{anh}},$$

where $c$ is a parameter that corresponds to the reversibility of polarization, which depends on the surface energy of the domain walls, and the density of pinning sites.

The anhysteretic polarization is shaped like the Langevin equation, which is suitable for modelling polycrystalline thin films like HZO. The maximum anhysteretic polarization is defined as a saturation polarization parameter $P_s$, which corresponds to an ideal case with no losses. The input for the Langevin equation is the effective displacement field, $D_{\text{eff}}$, felt by individual domains, which includes the contribution of coupling to neighboring domains with the inter-domain coupling parameter $\alpha$ to the simple displacement field $D$ as,

$$D_{\text{eff}} = D + \alpha P_{\text{irr}}$$

This effective field is divided by a loss contribution from the density of formed domain walls $a$, which leads to the equation,

$$P_{\text{anh}} = P_s\left[\coth\left(\frac{D_{\text{eff}}}{a}\right) - \frac{a}{D_{\text{eff}}}\right]$$

Irreversible polarization includes a parameter describing the effective energy to break pinning sites, $k$, and is dependent on the previous polarization contributions, forming a differential equation,

$$dP_{\text{irr}} = \frac{P_{\text{anh}} - P_{\text{irr}}}{\delta k - \alpha(P_{\text{anh}} - P_{\text{irr}})} dD$$

where $\delta = \text{sign}\left(\frac{dD}{dt}\right)$.

The total current in a FE capacitor is a sum of currents caused by the polarization switching, a constant capacitance term, and a leakage term, $I_{leak}$, here described with a trap limited thermal emission through the Simmons formula from [12], totaling to

$$I_{\text{tot}} = A\frac{dP}{dt} + C\frac{dV}{dt} + I_{\text{leak}}$$

where $A$ is the area of the devices and $C = \epsilon_0 \epsilon_r A/d$ is the linear capacitance, which is calculated from the effective dielectric permittivity $\epsilon_{r,eff}$ and layer thickness $d$. The equation governing the leakage current is

$$I_{leak} = 2q\mu E\sqrt{\left(\frac{2\pi m^* m_e k_B T}{h^2}\right)^3} \exp\left(-\frac{q(\phi_i - \sqrt{qE/4\pi\epsilon_{r,eff}\epsilon_0})}{k_B T}\right)$$

where $q, m_e, k_B, T, h, \epsilon_0$ are the constants of electron charge, electron mass, Boltzmann constant, temperature, Planck's constant and vacuum permittivity. Model parameters $m^*, \mu, \epsilon_{opt}, \phi_i$ correspond to effective electron mass, mobility, and optical permittivity in the insulator, as well as the barrier height. Due to asymmetric device stack, the value $i$ corresponds to the barriers in positive and negative directions with the indexes $A$ and $B$, respectively. This leakage mechanism was used since it includes both bulk- and electrode limited conduction mechanisms [12].

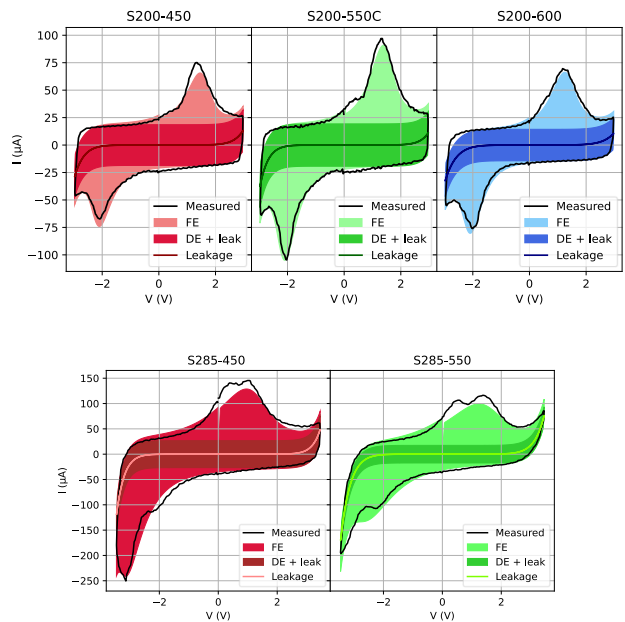

Fig. 3. Measured and simulated dynamic $I - V$ curves with emphasis on contributing parts to the current.

The simulated current shows excellent agreement with the experimental data, as is seen in Fig. 3, where the different current contributions are indicated with colors. Since the total current obtained corresponds well to the experimental results, the model parameters reveal details on the effect of different microcrystalline properties on the polarization switching as well as the magnitude of leakage and dielectric current contributions. Due to the device-to-device variability, we consider the average of obtained parameter values from the curves in Figures 1 and 2. Notably the Jiles-Atherton model cannot directly emulate the current peak splitting seen in S285 samples, and the current peak on the negative side, and hence, separating leakage and polarization switching peaks is not fully reliable. However, the voltage and frequency dependent properties of the leakage current can be used to obtain a more detailed description.

The physically interpretable average modelling parameters that can reconstruct polarization and different current contributions such as in Fig. 3. are shown in Table I for the Jiles-Atherton parameters, and for the leakage and dielectric permittivity in Table II.

TABLE I. AVERAGE MODELLING PARAMETERS FOR POLARIZATION

|  | $a$ | $\alpha$ | $k$ | $c$ | $P_s$ |
|---|---|---|---|---|---|
| **S200-450** | 0.0632 | 3.67 | 0.473 | 0.894 | 0.0882 |
| **S200-550** | 0.0863 | 2.58 | 0.467 | 0.835 | 0.126 |
| **S200-600** | 0.0356 | 1.71 | 0.587 | 0.624 | 0.161 |
| **S285-450** | 0.0615 | 0.909 | 0.567 | 0.879 | 0.259 |
| **S285-550** | 0.0584 | 0.435 | 0.591 | 0.803 | 0.358 |

Notable differences between the higher temperature ALD grown samples is the increased $P_s$, which indicates that more ferroelectric phase is present in the samples, when compared to the S200 series. Notably, the inter-domain coupling parameter is significantly smaller in the S285 series when compared to S200 series, which is in accordance with a larger local field opposing polarization switching, leading into higher coercive fields in the S285 samples when compared to S200 samples. The parameter $k$ describes the loss due to domain de-pinning, and it is clear from the table that it increases due to high temperature growth or annealing. This indicates that a higher number of mobile pinning sites are created due to high temperature thermal treatment. The polarization reversibility parameter describes immobile trap sites that cause reversible domain wall bending, and notably it decreases as the RTA or growth temperature is increased. Therefore, the results point that a higher temperature during fabrication leads to the trap sites which can become more easily mobile during field cycling, while a lower temperature fabrication also creates these traps, however the traps remain more immobile during field cycling.

TABLE II. AVERAGE MODELLING PARAMETERS FOR LEAKAGE

|  | $m^*$ | $\phi_A$ | $\phi_B$ | $\epsilon_{opt}$ | $\mu$ | $\epsilon_{r,eff}$ |
|---|---|---|---|---|---|---|
| **S200-450** | 2.40 | 1.199 | 1.217 | 2.65 | 8.10 | 39.8 |
| **S200-550** | 3.68 | 1.198 | 1.224 | 3.07 | 7.52 | 31.8 |
| **S200-600** | 3.84 | 1.193 | 1.456 | 4.90 | 7.47 | 31.5 |
| **S285-450** | 0.68 | 1.177 | 1.203 | 3.06 | 8.49 | 51.9 |
| **S285-550** | 1.06 | 1.128 | 1.147 | 9.62 | 9.75 | 43.2 |

From the values in Table II, it is visible that the effective relative permittivity of the dielectric and ferroelectric layers is higher in the S285 series than S200 series, indicating that higher temperature in the ALD deposition leads to a larger dielectric capacitance of the material. A lower potential barrier indicates that the leakage on the corresponding side starts increasing from a lower voltage. Notably, the positive potential barrier is almost constant in the S200 series, indicating that the temperature of the RTA process does not affect the leakage on the positive side as much as it does on the negative side. In the S285 series lower potential barriers are found to be a cause for the increase in the leakage. In S200 series on the negative side, on an average there is an increase in the barrier potential with the annealing temperature. The barrier potential depends not only on the bulk of the insulator, but the interface quality and the growth of unintentional dead layers. From the insulator properties, it is visible that clear changes happen in all the parameters, further indicating that the leakage properties are not only governed by the electrodes and dead layers, but also the ferroelectric material quality, where trap sites like oxygen vacancies influence the electron movement inside the ferroelectric material.

From the results, it can be clearly seen that different growth and annealing conditions can modify the sizes of the HZO crystalline grains. Formation of larger crystallites due to higher temperature growth or anneal result in larger $P_r$, leading to better D2D performance uniformity, however, with a higher leakage current penalty, verified by both modelling and experiments. The leakage current density in the large area capacitors (µA/cm$^2$) are in the range of $10^2$ to $10^3$ at 3V for S200-450 and S200-550 respectively, while it is ~$5 \times 10^3$ for S200-600, over $10^4$ at 3V for S285-450 and S285-550, exhibiting approximately two orders of magnitude increase in leakage current density for the samples grown at 285 $^0$C compared to 200 $^0$C for the same annealing temperatures. It has been shown in earlier studies that lower amount of the amorphous phase and the formation of larger crystallites in HZO can significantly increase leakage currents as both the effects can facilitate diffusion and electric field-driven drift of charge carriers along the crystalline grain boundaries, leading to early dielectric breakdown. [13]. Our modelling work involving both ferroelectric and leakage current model, together with the experimental validation throws new light on the effect that both higher deposition and annealing temperature has similar effects on the microcrystalline properties of HZO of forming large crystals, improving $P_r$ and uniformity at the cost of more mobile traps and higher leakage current paths, causing early dielectric breakdown. Larger crystallites can enable single-grain FeFETs in highly scaled technology nodes with low D2D variability [14]. High temperature and longer annealing time is a known way to get higher crystallite size, higher $P_r$ and better uniformity [5] while lower annealing temperature, necessary for BEOL process is responsible for inhomogeneous crystallization leading to large D2D variation. In a recent work [15], it is shown that the pinning loss and polarization reversibility parameters from ferroelectric Jiles Atherton model ($k$ and $c$) have a one-to-one correlation, since both are related to the number of microcrystalline defects that are likely caused by the inhomogeneous crystallization process within the CMOS BEOL thermal budget restriction. We show comprehensively that growth temperature of 280 $^0$C and annealing temperature of 450 $^0$C can already lead to larger crystallite size with lower non-uniformity of $P_r$ and coercive voltage making them compatible with ultra-scaled CMOS BEOL FeRAM or FeFET integration, where absence of grain boundaries can alleviate or at least minimize the leakage current issue. Also, a more rigorous optimization over the thermal parameter space is important for finding the sweet spot where high enough polarization and low leakage current is feasible.

## CONCLUSION

In conclusion, we report on thermal engineering of $Hf_{0.5}Zr_{0.5}O_2$ films that can lead to CMOS BEOL compatible ferroelectric components with high remnant polarization, yield, reliability and improved device-to-device performance uniformity even within BEOL compatible processing conditions paving their way for ultra-energy efficient memory and neuromorphic hardware. Based on stack composition and fabrication processes, fine tuning of these parameters might be necessary. However, the current work verifies that with proper device engineering, large-scale circuit implementation of back-end compatible ferroelectric devices are possible.

## ACKNOWLEDGMENT

The authors acknowledge financial support from Research Council of Finland through projects AI4AI (no. 352860), Ferrari (359047) and Business Finland and European Commission through KDT-JU project ARCTIC (Grant agreement No.

**REFERENCES**

[1] Y. Wang, "Memory for Data-Centric Computing: A Technology Perspective," 2020 International Symposium on VLSI Technology, Systems and Applications (VLSI-TSA), Hsinchu, Taiwan, 2020, pp. 21-21, doi: 10.1109/VLSI-TSA48913.2020.9203724.

[2] S Majumdar, "Back-End CMOS Compatible and Flexible Ferroelectric Memories for Neuromorphic Computing and Adaptive Sensing", Adv. Intell. Syst. vol. 4, pp. 2100175, 2022.

[3] U. Schroeder et al., "The fundamentals and applications of ferroelectric $HfO_2$", Nat. Rev. Mater. vol. 7, pp. 653–669, 2022.

[4] H. Bohuslavskyi, K. Grigoras, M. Ribeiro, M. Prunnila, S. Majumdar, "Ferroelectric $Hf_{0.5}Zr_{0.5}O_2$ for Analog Memory and In‑Memory Computing Applications Down to Deep Cryogenic Temperatures", Adv. Electron. Mater., vol. 10, pp. 2300879, 2024.

[5] S. De et al., "Uniform Crystal Formation and Electrical Variability Reduction in Hafnium-Oxide-Based Ferroelectric Memory by Thermal Engineering", ACS Appl. Electron. Mater. vol. 3(2), pp. 619–628, 2021.

[6] X. Li et al., "Designing high endurance $Hf_{0.5}Zr_{0.5}O_2$ capacitors through engineered recovery from fatigue for non-volatile ferroelectric memory and neuromorphic hardware," arXiv preprint arXiv:2409.00635. https://arxiv.org/abs/2409.00635.

[7] D. H. Lee et al., "Effect of residual impurities on polarization switching kinetics in atomic-layer-deposited ferroelectric $Hf_{0.5}Zr_{0.5}O_2$ thin films," Acta Materialia, vol. 222, pp. 117405, 2022.

[8] N. Aspiotis et al., "Large-area synthesis of high electrical performance $MoS_2$ by a commercially scalable atomic layer deposition process", npj 2D Mater. Appl. vol.7, pp. 18, 2023.

[9] S. Asapu et al., "Large remnant polarization and great reliability characteristics in W/HZO/W ferroelectric capacitors." Frontiers in Materials, vol. 9, pp. 969188, 2022.

[10] H. -G. Kim, et al. "Effect of process temperature on density and electrical characteristics of $Hf_{0.5}Zr0_{.5}O_2$ thin films prepared by plasma-enhanced atomic layer deposition." Nanomater. vol. 12, pp. 548, 2022.

[11] E. Paasio, M. Prunnila and S. Majumdar, "Modelling Ferroelectric Hysteresis of HZO Capacitor with Jiles-Atherton Model for Non-Volatile Memory Applications," 2023 IEEE 12th Non-Volatile Memory Systems and Applications Symposium (NVMSA), Niigata, Japan, 2023, pp. 1-2, doi: 10.1109/NVMSA58981.2023.00019.

[12] F.-C. Chiu, "A Review on Conduction Mechanisms in Dielectric Films," Advances in Materials Science and Engineering, vol. 2014, pp. 1–18, 2014, doi: 10.1155/2014/578168.

[13] K. McKenna, A. Shluger, V. Iglesias, M. Porti, M. Nafría, M. Lanza, G. Bersuker, "Grain boundary mediated leakage current in polycrystalline $HfO_2$ films", Microelectronic Engineering Vol. 88, Issue 7, pp. 1272-1275, 2011.

[14] M. Lederer, D. Lehninger, T. Ali, T. Kämpfe, "Review on the Microstructure of Ferroelectric Hafnium Oxides", Physica Status Solidi, RRL vol.16, issue10, pp. 2200168, 2022.

[15] E. Paasio, R. Ranta, S. Majumdar, "Universal Model for Ferroelectric Capacitors Operating Down to Deep Cryogenic Temperatures", arXiv preprint arXiv:2410.09131. https://arxiv.org/abs/2410.09131